\documentclass[
superscriptaddress,
twocolumn,
amsmath,amssymb,
aps,
prb,
floatfix,
showkeys,
reprint,
]{revtex4-2}

\usepackage{graphicx}
\usepackage{dcolumn}
\usepackage{bm}
\usepackage{amsmath}
\usepackage{amssymb}
\usepackage{multirow}
\usepackage{xcolor}
\usepackage{booktabs}
\usepackage{tabularx}
\usepackage[english]{babel}
\usepackage[utf8]{inputenc}
\usepackage{enumerate}
\usepackage{newunicodechar}
\newunicodechar{ﬁ}{fi}
\newunicodechar{ﬀ}{ff}
\usepackage[colorlinks, citecolor=blue, linkcolor=blue]{hyperref}

\newcommand{\Soum}[1]{{ \color{black} #1}}
\newcommand{\Stef}[1]{{ \color{black} #1}}

\begin{document}
\title{Spontaneous square vs.~hexagonal nanoscale skyrmion lattices in Fe/Ir(111)}

\author{Mara Gutzeit}
\affiliation{Institute of Theoretical Physics and Astrophysics, University of Kiel, Leibnizstrasse 15, 24098 Kiel, Germany}

\author{Tim Drevelow}
\affiliation{Institute of Theoretical Physics and Astrophysics, University of Kiel, Leibnizstrasse 15, 24098 Kiel, Germany}

\author{Moritz A. Goerzen}
\affiliation{Institute of Theoretical Physics and Astrophysics, University of Kiel, Leibnizstrasse 15, 24098 Kiel, Germany}

\author{Soumyajyoti Haldar}
\email[Corresponding author: ]{haldar@physik.uni-kiel.de}
\affiliation{Institute of Theoretical Physics and Astrophysics, University of Kiel, Leibnizstrasse 15, 24098 Kiel, Germany}

\author{Stefan Heinze}
\affiliation{Institute of Theoretical Physics and Astrophysics, University of Kiel, Leibnizstrasse 15, 24098 Kiel, Germany}
\affiliation{Kiel Nano, Surface, and Interface Science (KiNSIS), University of Kiel, Germany}

\date{\today}

\begin{abstract}
We study the emergence of spontaneous skyrmion lattices in an Fe monolayer in fcc and
hcp stacking on the Ir(111) surface using density functional theory (DFT).  
For fcc-Fe/Ir(111) we find the well-known square nanoskyrmion lattice.  
However, for hcp-Fe/Ir(111) the hexagonal 
skyrmion lattice previously proposed based on experiments
is energetically unfavorable with respect to a 
hexagonal multi-Q state with nearly collinear magnetic moments.
By mapping our DFT calculations to an atomistic 
spin model we demonstrate that the interplay of pair-wise exchange, higher-order exchange, and Dzyaloshinskii-Moriya interaction is decisive for the symmetry and collinearity of the obtained
spin lattice. 
\end{abstract}

\maketitle

Spintronics using non-collinear spin textures has been a rapidly growing field since the experimental
discovery of magnetic skyrmion lattices \cite{Muhlbauer2009,Yu2010,Heinze2011}.
Skyrmions
have been observed in a large variety of magnetic materials even at
room temperature \cite{Moreau2016,Boulle2016,Woo2016} and
numerous potential applications of magnetic skyrmions \cite{Bogdanov1989}
are currently being explored \cite{Fert2017,Back2020}. 
The origin of skyrmions is the 
Dzyaloshinskii-Moriya interaction (DMI) \cite{Dzyaloshinskii1957,Moriya1960}
which occurs due to spin-orbit coupling 
in systems with broken inversion symmetry and favors canted spin structures
with a unique rotational sense \cite{Bode2007}. It has been shown that frustrated
or higher-order exchange can stabilize nanoscale topological spin structures
as well \cite{Okubo2012,Leonov2015,Malottki2017,Paul2020}. Higher-order exchange
interactions such as the four-spin interaction can also lead to 
multi-Q states which are a superposition of single-Q (spin spiral) 
states \cite{Kurz2001,Kroenlein2018,Romming2018,Spethmann2020,Li2020,Haldar2021,Gutzeit2021}.

The formation of a spontaneous nanoskyrmion lattice in an Fe monolayer (ML) grown in fcc
stacking on the Ir(111) surface, denoted as fcc-Fe/Ir(111), %
is caused by the interplay of DMI and the four-spin exchange interaction \cite{Heinze2011}. Recently, it has been demonstrated that there are actually
two types of four-spin exchange terms \cite{Hoffmann2020}.
This can explain the collinear %
up-up-down-down ($uudd$) ground state
-- a 2Q state formed from the superposition of two $90^\circ$ spin spirals --
of an Fe ML on Rh(111) \cite{Kroenlein2018} and the slightly canted 
$uudd$ state found in Rh/Fe/Ir(111) \cite{Romming2018}.
The newly proposed three-site four-spin interaction \cite{Hoffmann2020}
can also lead to two-dimensional nearly
collinear multi-Q states as observed in an Fe ML in hcp stacking on 
Rh/Ir(111) \cite{Gutzeit2022}. In contrast to the square nanoskyrmion 
lattice of fcc-Fe/Ir(111) \cite{Heinze2011} these collinear multi-Q states 
possess a hexagonal unit cell. For hcp-Fe/Ir(111) 
a hexagonal magnetic state
has been observed using
spin-polarized scanning tunneling microscopy (SP-STM) and interpreted based on
the experimental data as a nanoskyrmion lattice \cite{Bergmann2015}. 
However, a first-principles study of hcp-Fe/Ir(111) has been missing so far.

Here, we investigate the Fe monolayer in both fcc and hcp stacking on Ir(111) by means of first-principles calculations based on DFT. 
We find that a non-collinear square nanoskyrmion lattice is the magnetic ground
state of fcc-Fe/Ir(111) in accordance with previous work~\cite{Heinze2011}.
Surprisingly,
a nearly collinear hexagonal multi-Q state is 
revealed to be lowest in energy for hcp-Fe/Ir(111) in contrast to the hexagonal 
non-collinear skyrmion lattice
previously proposed based on spin-polarized scanning tunneling microscopy
measurements~\cite{Bergmann2015}. In order to understand the
stabilization mechanisms of spontaneous skyrmion lattices
with different symmetry and degree of non-collinearity
we map total DFT energies of a variety of complex magnetic structures to an atomistic spin model. We reveal that in Fe/Ir(111) the interplay of 
pairwise Heisenberg exchange, Dzyaloshinskii-Moriya interaction (DMI), and 
higher-order exchange interactions favors a high degree of non-collinearity 
in square nanoskyrmion lattices while it leads to nearly collinear spin
alignments in hexagonal spin lattices. 

\begin{figure}[htbp!]
	\centering
	\includegraphics[width=0.85\linewidth]{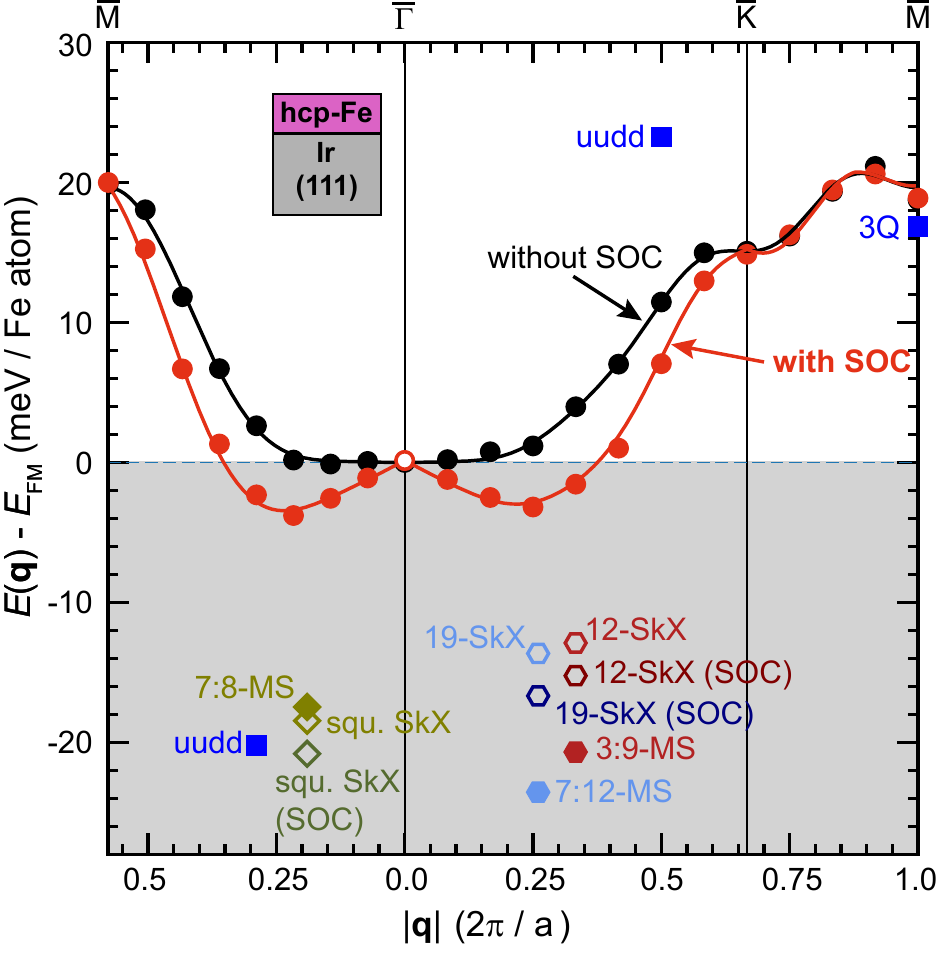}
	\caption{Energy dispersion E(\textbf{q}) of flat cycloidal spin spirals calculated via DFT along the two high symmetry directions of the two-dimensional Brillouin zone for hcp-Fe/Ir(111). Black (red) circles represent total DFT energies without (with) spin-orbit coupling (SOC), while black (red) lines show a fit to the Heisenberg model neglecting (including) DMI. The DFT total energies of several multi-Q states discussed in the text are denoted at the \textbf{q} values of their constituting single-Q states.}
	\label{fig:SSDispersion}
\end{figure} 

We start by discussing the energy dispersion $E(\textbf{q})$ of flat spin spirals for both stackings of the Fe ML on Ir(111)
[Fig.~\ref{fig:SSDispersion} and Suppl.~Fig.~S1~\cite{supplmat}]
obtained via DFT calculations using the {\tt FLEUR} code \cite{FLEUR}
(see Supplemental Material~\cite{supplmat} for computational details). 
Spin spirals are characterized by a $\mathbf{q}$ vector from the irreducible part of the two-dimensional hexagonal Brillouin zone (2D BZ). They represent the general solution
of the Heisenberg model on a periodic lattice and thus allow 
to scan a large part of the magnetic phase space. 
For a spin spiral with a specific wave vector $\mathbf{q}$ the direction of a magnetic moment at lattice site $\textbf{R}_i$ is given by $\textbf{M}_i$=$M(\cos(\mathbf{q}\cdot \mathbf{R}_i),\sin(\mathbf{q}\cdot \mathbf{R}_i),0)$, with $M$ denoting the size of the magnetic moment.  The calculated Fe magnetic moments are $M_{\text{Fe}}\approx 2.72$~$\mu_B$ for fcc-Fe/Ir(111) and $\approx 2.71$~$\mu_B$ for hcp-Fe/Ir(111), respectively, and are quite stable upon the variation of $\mathbf{q}$.

We find a large exchange frustration for both 
Fe stackings. For hcp-Fe/Ir(111) [Fig.~\ref{fig:SSDispersion}], the energy 
dispersion neglecting spin-orbit coupling is quite flat around the
ferromagnetic (FM) state at the $\overline{\Gamma}$ point. The row-wise 
antiferromagnetic (RW-AFM) state at the $\overline{\text{M}}$ point is 
by about \Soum{20} meV/Fe atom higher in energy than the FM state. 
For fcc-Fe/Ir(111) [Suppl.~Fig.~S1~\cite{supplmat}], a similar picture
emerges, however, the FM state represents a local energy maximum while spin spirals with periods
of $\lambda=1.9-1.7$~nm ($q =|\mathbf{q}|\approx 0.14-0.16 \times 2\pi/a$) along both high symmetry directions experience a small energy gain of 2 to 4 meV/Fe atom. 
Mapping the DFT results to the Heisenberg model of pair-wise exchange (black lines) reveals a small FM nearest-neighbor exchange constant which competes with AFM interactions of second- and third-nearest neighbors (see Table~\ref{Table1}).

\begin{table}[htb]
	\centering
	\caption{Heisenberg exchange constants for the first three nearest neighbors, $J_1$ to $J_3$, as extracted from the fit of the respective energy dispersion (without modification by \Soum{higher-order interaction} terms) and the higher-order exchange constants $B_1$, $K_1$ and $Y_1$ calculated using the pseudo-inverse method (see 
    Supplemental Material~\cite{supplmat} for details). All values are given in meV.}
	\label{Table1}
	\begin{ruledtabular}
		\begin{tabular}{l c c c c c c }
			System& $J_1$ &$J_2$ &$J_3$ & $B_1$& $K_1$ & $Y_1$ \\
			\colrule
            fcc-Fe/Ir(111)&5.46 &$-1.35$ &$-1.24$&$-1.97$ &$-2.22$ &2.47 \\
            hcp-Fe/Ir(111)& \Soum{2.52} & $\Soum{-0.24}$ & $\Soum{-1.03}$ &$\Soum{0.42}$ &$\Soum{-2.09}$ & \Soum{0.68} \\

			 \end{tabular} 
	\end{ruledtabular}
\end{table}	

The inclusion of spin-orbit coupling (SOC) in DFT calculations
[Fig.~\ref{fig:SSDispersion}] 
generates energy minima in hcp-Fe/Ir(111) for cycloidal spin spirals with 
periods of $\lambda=\Soum{1.4-1.1}
$~nm ($q =|\mathbf{q}|\approx \Soum{0.20-0.25}%
\times 2\pi/a$) with a depth of up to \Soum{4}
meV/Fe atom and further stabilizes the 
spin spiral minima in \Soum{fcc-Fe/Ir(111)}.
The energy contributions due to SOC for every spin spiral state 
\Soum{[Suppl. Fig. S2~\cite{supplmat}]} 
reveals the preference of a clockwise rotational sense of cycloidal spin spirals
for both Fe stackings with a nearest-neighbor DMI constant of about \Soum{1.6} meV 
(see Supplemental Material~\cite{supplmat} for all DMI constants).

To go beyond spin spiral (single-Q) states %
and search for the experimentally observed square and hexagonal spin lattices \cite{Heinze2011,Bergmann2015}, we have calculated the total DFT energies of several multi-Q states. We find that for both Fe stackings
the %
collinear $uudd$ (2Q) state along the $\overline{\Gamma \rm M}$ direction is much lower in energy than all spin spiral states [Fig.~\ref{fig:SSDispersion}
and Fig.~S1].
In contrast, the 
$uudd$-$\overline{\Gamma \rm K}$ state and the triple-Q state are on the order of \Soum{20}
to 40 meV/Fe atom above the FM state.
Within the Heisenberg model of pair-wise exchange a multi-Q state and its corresponding 
spin spiral (1Q) state are energetically degenerate. Therefore, the large energy differences
obtained in our DFT calculations indicate significant higher-order exchange contributions.

\begin{figure}[htbp!]
	\centering
	\includegraphics[width=0.85\linewidth]{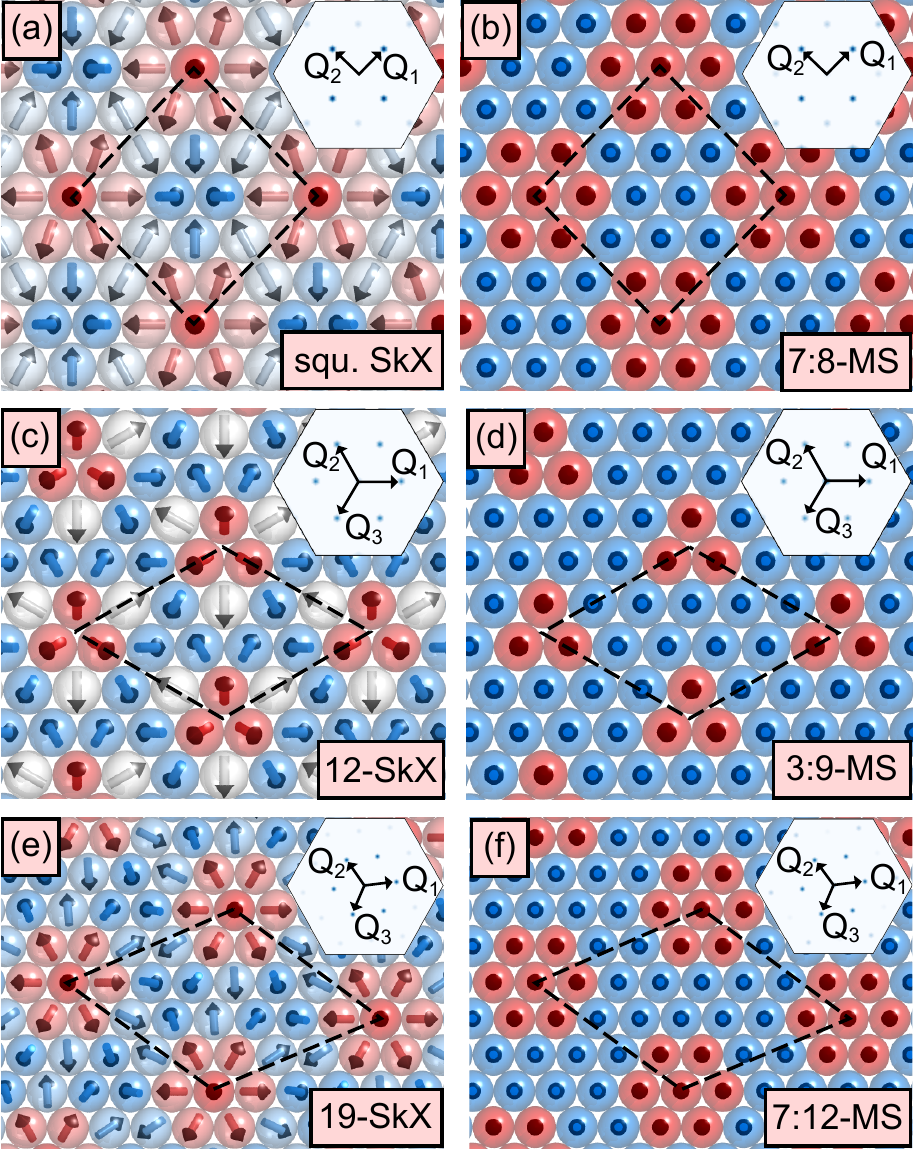}
	\caption{Sketches of selected multi-Q states in Fe/Ir(111) with their respective unit cells indicated by dashed lines. (a) the square nanoskyrmion lattice observed in fcc-Fe/Ir(111)~\cite{Heinze2011} and (b) the corresponding
	collinear 7:8-MS. 
	(c) the hexagonal 12-atomic skyrmion lattice proposed as the magnetic ground state of hcp-Fe/Ir(111)~\cite{Bergmann2015} and (d) the corresponding
	collinear 3:9-MS. (e) the hexagonal 19-atomic skyrmion lattice 
	and (f) the corresponding collinear 7:12-MS.
	Spheres illustrate Fe atoms and arrows the direction of their magnetic moments in which red (blue) denotes the up (down) out-of-plane component. Insets
	show the Fourier transform of the spin structures in the hexagonal 
	Brillouin zone and the two (three) $\mathbf{Q}$ vectors from which the 
	square (hexagonal) skyrmion lattices are constructed.}
	\label{fig:Magneticstates}
\end{figure}

We have further considered the
square nanoskyrmion lattice (squ SkX) observed in fcc-Fe/Ir(111) \cite{Heinze2011} that is constructed from a superposition of two $\textbf{q}$ vectors enclosing an angle of about 90$^{\circ}$ [Fig.~\ref{fig:Magneticstates}(a)]. 
In the scalar-relativistic calculation
this state %
\Stef{is energetically slightly above}
the $uudd$ state [Fig.~\ref{fig:SSDispersion}]. However, the inclusion of SOC
significantly lower its energy by 
\Stef{by about 2 meV/Fe atom for both stackings.} 
By projecting the magnetic moments of the %
squ SkX onto the $z$ axis perpendicular to the film plane we obtain a 
fully collinear state with seven moments pointing in one and eight pointing into the opposite direction, the 7:8-mosaic state (MS) [Fig.~\ref{fig:Magneticstates}(b)] \cite{vonBergmann2006}. In both 
systems, this state is energetically clearly unfavorable compared to the squ SkX
[Fig.~\ref{fig:SSDispersion} and Fig.~S1]. 

Hexagonal skyrmion lattices can be constructed
from the superposition of three $\textbf{q}$ vectors of equal lengths and with angles 
of 120$^{\circ}$ between them \cite{Gutzeit2022}. If one chooses the $\textbf{q}$ vectors 
along the symmetry-equivalent $\overline{\Gamma \rm K}$ directions and with a period
of three nearest-neighbor distances one obtains a hexagonal 12 atom SkX [Fig.~\ref{fig:Magneticstates}(c)] 
previously 
proposed as the magnetic ground state of hcp-Fe/Ir(111)~\cite{Bergmann2015}. The
total DFT energy of the hex 12-SkX is lower than the spin spiral minimum
\footnote{Note, that the 12-SkX state can also occur as an {\it on-top-state} 
in which the point of constructive interference of the three 
spin spirals is placed on top of a Fe atom \cite{Bergmann2015}. In contrast, for the 12-SkX state of Fig.~2(c) this point is
on a hollow site. Within our DFT calculations, the {\it on-top} and {\it hollow}
12-SkX states are energetically nearly degenerate (see Supplemental Material \cite{supplmat}).}
but significantly above the squ SkX [Fig.~\ref{fig:SSDispersion}]. Surprisingly,
the collinear analogue, the 3:9-MS [Fig.~\ref{fig:Magneticstates}(d)], gains energy with respect to the corresponding hex 12-SkX for both Fe stackings. This effect is even 
more prominent in hcp-Fe/Ir(111) [Fig.~\ref{fig:SSDispersion}] and brings the energy of the 3:9-MS
\Stef{below} the $uudd$-$\overline{\Gamma \rm M}$ state and 
\Stef{to the same energy as} the square SkX. 

A similar scenario emerges for the hexagonal 
19-SkX vs.~the collinear 7:12-MS 
which are constructed in an analogous way with shorter \Soum{and slightly rotated} $\textbf{q}$ vectors [Figs.~\ref{fig:Magneticstates}(e,f)]. 
The collinear 7:12-MS, which has previously been observed in 
hcp-Fe/Rh/Rh/Ir(111) \cite{Gutzeit2022}, is energetically lower than the 
non-collinear hexagonal 19-SkX [Fig.~\ref{fig:SSDispersion}]. 
For hcp-Fe/Ir(111) the 7:12-MS state is energetically even lower than the 
\Stef{3:9-MS}
and the square SkX [Fig.~\ref{fig:SSDispersion}].

Note, that experimentally a hexagonal magnetic structure with 12 atoms per unit cell,
consistent with the 3:9-MS or the 12-SkX, has been observed in hcp-Fe/Ir(111) using 
SP-STM \cite{Bergmann2015}, while in our DFT calculation the 7:12-MS is slightly lower
in total energy [Fig.~\ref{fig:SSDispersion}]. Nevertheless, 
the trend favoring a {\it hexagonal collinear} rather than a {\it square non-collinear} 
spin lattice for hcp-Fe stacking can clearly be recognized from our DFT calculation. 
In contrast, the non-collinear square SkX is the energetically lowest state 
for fcc-Fe/Ir(111) in accordance with previous work [Suppl.~Fig.~S1~\cite{supplmat}]~\cite{Heinze2011}.

In order to explain the microscopic origin of a collinear hexagonal mosaic state in hcp-Fe/Ir(111) 
vs.~a non-collinear square nanoskyrmion lattice in fcc-Fe/Ir(111) we consider an atomistic spin model
given by
\begin{equation}
\label{eq:Hamiltonian}
\begin{gathered}
 H=  -\sum_{ij} J_{ij}(\mathbf{m}_i \cdot \mathbf{m}_j) -\sum\limits_{ij} \mathbf{D}_{ij} ( \mathbf{m}_i \times \mathbf{m}_j)\\ -\sum\limits_{i} K_{u}(m_i^z)^2  -\sum_{<ij>} B_1(\mathbf{m}_i \cdot \mathbf{m}_j)^2\\
 -\sum_{<ijkl>}K_1[(\mathbf{m}_i \cdot \mathbf{m}_j)(\mathbf{m}_k \cdot \mathbf{m}_l)+(\mathbf{m}_i \cdot \mathbf{m}_l)(\mathbf{m}_j \cdot \mathbf{m}_k)\\
 -(\mathbf{m}_i \cdot \mathbf{m}_k)(\mathbf{m}_j \cdot \mathbf{m}_l)]\\
 -\sum_{<ijk>}Y_1[(\mathbf{m}_i \cdot \mathbf{m}_j)(\mathbf{m}_j \cdot \mathbf{m}_k)+(\mathbf{m}_j \cdot \mathbf{m}_i)(\mathbf{m}_i \cdot \mathbf{m}_k)+\\
 (\mathbf{m}_i \cdot \mathbf{m}_k)(\mathbf{m}_k \cdot \mathbf{m}_j)]\text{.}
 \end{gathered}
 \end{equation}
Here, $\mathbf{m}_i=\mathbf{M}_i/M$%
denotes the normalized magnetic moment at lattice site $i$, $J_{ij}$ the Heisenberg pair-wise exchange constants,  $\mathbf{D}_{ij}$ the vectors of the DMI and $K_u$ the uniaxial magnetocrystalline anisotropy constant. The last three terms are the higher-order exchange interactions arising from fourth order perturbation theory from a multi-band Hubbard model~\cite{Hoffmann2020}. 
We restrict ourselves to the nearest-neighbor approximation as suggested in 
Ref.~\cite{Hoffmann2020} and the coupling strengths 
are $B_1$, $K_1$, and $Y_1$ for the biquadratic, four-site four spin, and
three-site four spin interaction, respectively (Table \ref{Table1}).
All interaction constants of the atomistic spin model, given by
Eq.~(\ref{eq:Hamiltonian}), were determined from DFT total energies (see Supplemental Material~\cite{supplmat} for details).

\begin{figure}[htbp!]
	\centering
\includegraphics[width=0.85\linewidth]{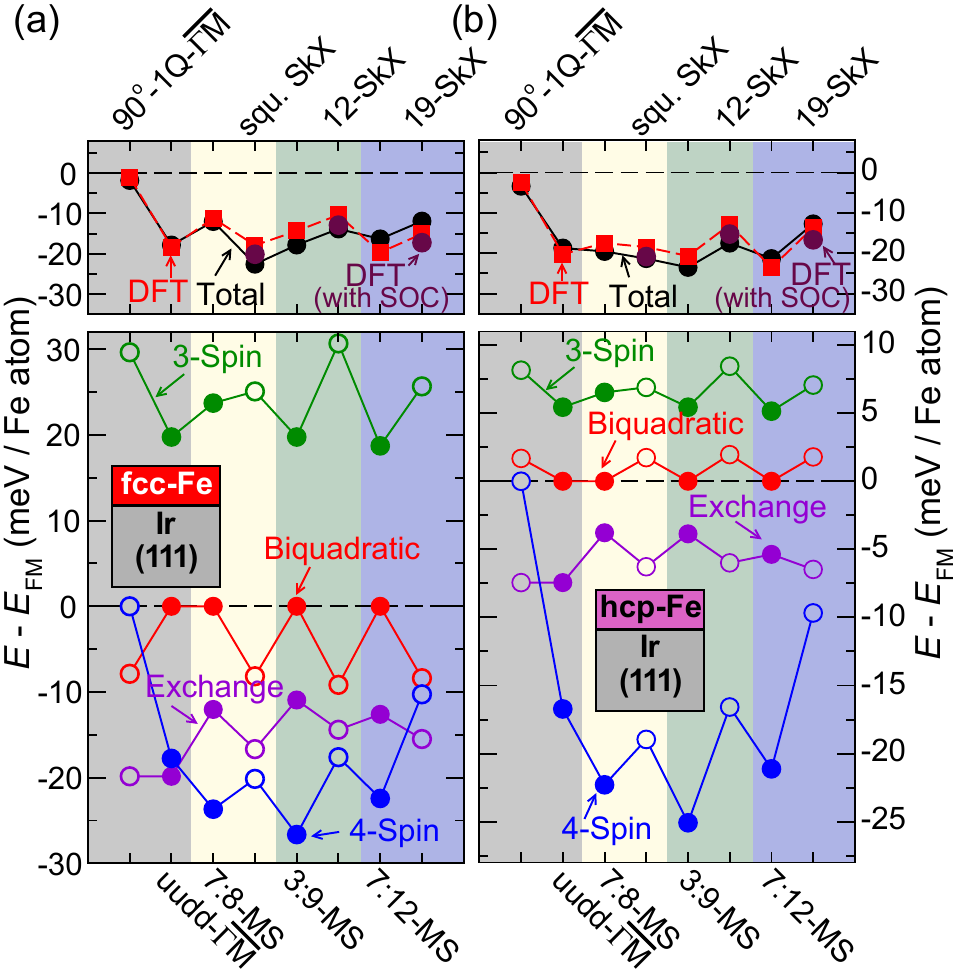}
	\caption{Spin model vs.~DFT total energies for selected magnetic states in (a) fcc-Fe/Ir(111) and (b) hcp-Fe/Ir(111). Upper panels show DFT total energies (red squares) and energies obtained via the atomistic spin model (black circles) using DFT parameters for the magnetic interactions (see Supplemental Material for all values).
	The top axis denotes the non-collinear spin states and the
	lower axis 
	the corresponding collinear states. 
	Lower panels show the decomposition of the energy into 
	the contributions from the Heisenberg exchange, the biquadratic
	interaction, %
	the three-site four spin (3-Spin) and the four-site four spin interaction (4-Spin). Filled (open) circles represent collinear (non-collinear) states.
    The lines connecting the data points serve as a guide to the eye.
    \Stef{Note that the spin model total energies include the DMI contribution not shown in the decomposition.}}
	\label{fig:Spinmodel}
\end{figure}

Fig.~\ref{fig:Spinmodel} shows the energy of the energetically lowest states of Fe/Ir(111) computed from the atomistic spin model
\footnote{Note, that the energies of all spin spiral (1Q) states in the spin model are equal to the DFT values.}.
The trend of DFT energies for the multi-Q states is captured 
by the atomistic spin 
model 
(top panels of Fig.~\ref{fig:Spinmodel}). In particular, 
the collinear 3:9-MS and 7:12-MS are energetically
preferred over the hexagonal 12-SkX and 19-SkX, respectively. In contrast, the non-collinear square SkX is 
lower in energy than the collinear 7:8-MS in agreement with DFT. 
The model also correctly predicts the square SkX as the magnetic ground state of fcc-Fe/Ir(111)
\Stef{and in the case of hcp-Fe/Ir(111) the hexagonal 3:9-MS state is the
state of lowest energy.}

\begin{figure}[htbp!]
	\centering
	\includegraphics[width=1\linewidth]{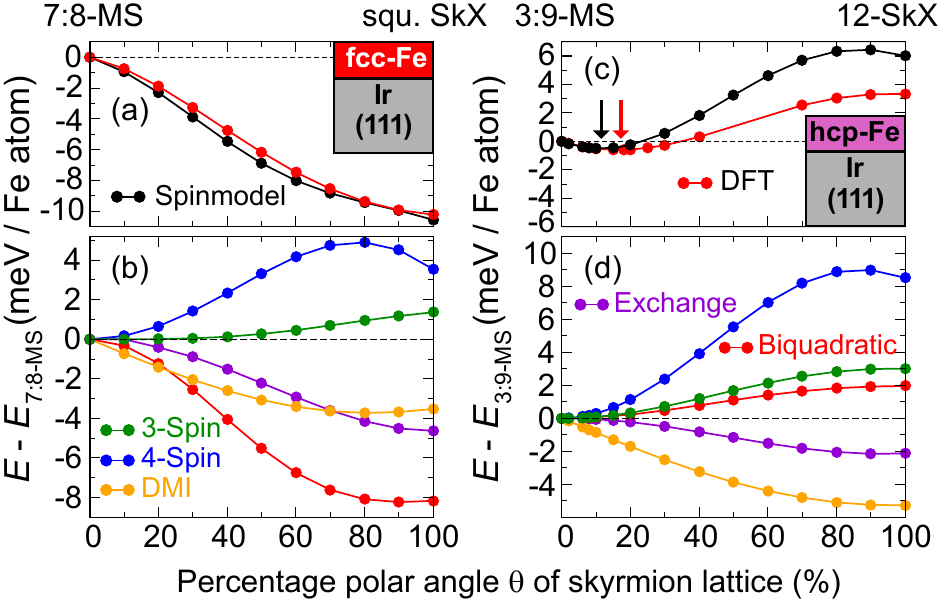}
	\caption{Total energies of magnetic states along the geodesic path from (a) the square 7:8-MS 
    to the square SkX for fcc-Fe/Ir(111)
	and (c) the hexagonal 3:9-MS to the hexagonal 12-SkX for hcp-Fe/Ir(111). 
	Red and black data points are obtained via DFT including the effect of SOC and via the atomistic spin model, respectively. (b) and (d) show the energy resolved contributions from the single magnetic interactions obtained in the atomistic spin model. 
	The relative polar angle $\theta$ is defined for every magnetic
	moment in the unit cell as $\theta(x)$=$\theta_0+x(\theta_f - \theta_0)$ with $x \in [0,1]$ where the value $x=0$ is chosen for the collinear MS and $x=1$ for the non-collinear SkX in each case. $\theta_f$ refers to the final value of every magnetic moment in the respective SkX, whereas $\theta_0$ is set to 0$^{\circ}$ for upward pointing moments (180$^{\circ}$ for downward pointing moments). The lines connecting the data points serve as a guide to the eye.}
	\label{fig:squ_hex_geodesicpath}
\end{figure}

By decomposing the total energy from the spin model into the single interaction contributions one can study how the differences between square and hexagonal skyrmion lattices arise (lower panels in Fig.~\ref{fig:Spinmodel}). 
\Stef{The %
four-site four spin %
interaction ($K_1<0$)  
leads to a coupling of 1Q (spin spiral) states to multi-Q states}
as previously reported for fcc-Fe/Ir(111)~\cite{Heinze2011}. 
For example, the $uudd$-$\overline{\Gamma {\rm M}}$ state is preferred by about 20~meV/Fe atom with respect to the 
$90^\circ$-1Q-$\overline{\Gamma {\rm M}}$ spin spiral state [Fig.~\ref{fig:Spinmodel}] and a similar energy
gain is obtained for skyrmion lattices and mosaic states vs.~1Q states. \Stef{Note, that the three mosaic states gain the largest amount of 
energy by the four-site four spin interaction.}

The three-site four spin interaction with its positive sign ($Y_1>0$)
promotes collinear over non-collinear spin states with a considerable preference for hexagonal spin lattices. This effect has previously been reported to stabilize collinear hexagonal multi-Q states in Fe/Rh/Ir(111) \cite{Gutzeit2022}
\Stef{A decisive role is played by the biquadratic
term. Due to its sign, it favors 
non-collinear states over collinear states for 
fcc-Fe stacking 
($B_1<0$) and vice versa for hcp-Fe stacking ($B_1>0$). Thereby, the square SkX becomes the
energetically lowest state for fcc-Fe/Ir(111),
while the hexagonal 3:9-MS is favored for hcp-Fe/Ir(111).}

\Stef{To obtain a deeper insight into the question}
why the square spin lattice is non-collinear while the hexagonal spin lattices are collinear,
we have calculated via DFT the total energies of magnetic states in which the moments rotate continuously from the 7:8-MS to the square SkX and from the hexagonal 3:9-MS to the 12-SkX, respectively [Fig.~\ref{fig:squ_hex_geodesicpath}].

For the square spin lattice of fcc-Fe/Ir(111) [Fig.~\ref{fig:squ_hex_geodesicpath}(a)] the fully non-collinear skyrmion lattice clearly represents the %
energy minimum along the geodesic path in spin space. The atomistic spin model gives an excellent 
quantitative description of the DFT energies. 

In contrast, for the hexagonal 12-atomic state of hcp-Fe/Ir(111) 
the energy minimum is 
found close to the collinear 3:9-MS in both DFT and spin model
(see red/black arrow in Fig.~\ref{fig:squ_hex_geodesicpath}(c)).
In the DFT calculation this 18\% canted state gains 0.6 meV/Fe atom with respect to the collinear state. 
Note, that the deviation from the collinear MS state corresponds to only 7$^{\circ}$ to 16$^{\circ}$ of the magnetic moments from the $z$ direction. Hence the degree of non-collinearity is very small. The spin model shows a similar trend of energy 
vs.~canting, however, the quantitative deviations are larger than for the square
spin lattice [Fig.~\ref{fig:squ_hex_geodesicpath}(a)].

The different behavior of square vs.~hexagonal skyrmion lattices can be explained by the decomposition of the total energy from the spin model [Figs.~\ref{fig:squ_hex_geodesicpath}(b,d)]. 
In both cases the four-spin interactions favor the collinear MS, while the
pair-wise exchange and the DMI promote the non-collinear SkX. 
\Stef{Due to the opposite sign of the biquadratic term 
for the two Fe stackings, it favors the non-collinear SkX in 
fcc-Fe while it stabilizes the collinear MS for hcp-Fe.}
Quantitatively
the contributions of the competing interactions differ between square and hexagonal spin lattices.
For the square spin lattice [Fig.~\ref{fig:squ_hex_geodesicpath}(b)], the effect of four-spin interactions is weak 
\Stef{and the exchange, DMI, and biquadratic term
stabilize} %
the SkX. For the hexagonal spin lattice [Fig.~\ref{fig:squ_hex_geodesicpath}(d)], the \Stef{biquadratic and}
four-spin interactions dominate and a nearly collinear MS occurs. 
\footnote{\Stef{Note that $K_1$ is of similar magnitude for both Fe stackings while $B_1$ and $Y_1$ are much smaller for hcp-Fe
(cf.~Table \ref{Table1}).}}

In order to directly compare our prediction of a nearly collinear hexagonal 
multi-Q state
with previous experiments on hcp-Fe/Ir(111) \cite{Bergmann2015},
we have simulated SP-STM images 
(see Supplemental Material~\cite{supplmat}). %
We find that the SP-STM images are very similar for
the previously proposed non-collinear 12-SkX and the nearly collinear 3:9-MS state 
predicted in this work irrespective of the magnetization direction of the spin-polarized
tip.
This makes an unambiguous experimental distinction challenging.
However, a similar preference of a hexagonal nearly collinear 12:15-MS over the corresponding SkX has recently been 
found by DFT
for hcp-Fe/Rh/Ir(111) and confirmed by SP-STM experiments \cite{Gutzeit2022}.

In conclusion, we have demonstrated that the appearance of non-collinear 
skyrmion lattices vs.~nearly collinear multi-Q states in Fe monolayers on Ir(111) 
depends on the symmetry of the spin states due to the interplay of pair-wise 
exchange, higher-order exchange, and DMI. While the magnetic moments in the
nanoskyrmion lattice of fcc-Fe/Ir(111) are strongly canted,
we propose that 
the magnetic ground state of hcp-Fe/Ir(111) is a nearly collinear hexagonal 
multi-Q state.

We thank Andr\'e Kubetzka for providing the images of the
spin structures shown in Figure 2 and Gustav Bihlmayer for insightful discussions.
We gratefully acknowledge financial support from the Deutsche Forschungsgemeinschaft (DFG, German Research Foundation) via projects no.~418425860 and no.~462602351
(SPP2137 ``Skyrmionics") and
computing time provided by the North-German Supercomputing Alliance (HLRN).


\begin{thebibliography}{32}%
	\makeatletter
	\providecommand \@ifxundefined [1]{%
		\@ifx{#1\undefined}
	}%
	\providecommand \@ifnum [1]{%
		\ifnum #1\expandafter \@firstoftwo
		\else \expandafter \@secondoftwo
		\fi
	}%
	\providecommand \@ifx [1]{%
		\ifx #1\expandafter \@firstoftwo
		\else \expandafter \@secondoftwo
		\fi
	}%
	\providecommand \natexlab [1]{#1}%
	\providecommand \enquote  [1]{``#1''}%
	\providecommand \bibnamefont  [1]{#1}%
	\providecommand \bibfnamefont [1]{#1}%
	\providecommand \citenamefont [1]{#1}%
	\providecommand \href@noop [0]{\@secondoftwo}%
	\providecommand \href [0]{\begingroup \@sanitize@url \@href}%
	\providecommand \@href[1]{\@@startlink{#1}\@@href}%
	\providecommand \@@href[1]{\endgroup#1\@@endlink}%
	\providecommand \@sanitize@url [0]{\catcode `\\12\catcode `\$12\catcode
		`\&12\catcode `\#12\catcode `\^12\catcode `\_12\catcode `\%12\relax}%
	\providecommand \@@startlink[1]{}%
	\providecommand \@@endlink[0]{}%
	\providecommand \url  [0]{\begingroup\@sanitize@url \@url }%
	\providecommand \@url [1]{\endgroup\@href {#1}{\urlprefix }}%
	\providecommand \urlprefix  [0]{URL }%
	\providecommand \Eprint [0]{\href }%
	\providecommand \doibase [0]{https://doi.org/}%
	\providecommand \selectlanguage [0]{\@gobble}%
	\providecommand \bibinfo  [0]{\@secondoftwo}%
	\providecommand \bibfield  [0]{\@secondoftwo}%
	\providecommand \translation [1]{[#1]}%
	\providecommand \BibitemOpen [0]{}%
	\providecommand \bibitemStop [0]{}%
	\providecommand \bibitemNoStop [0]{.\EOS\space}%
	\providecommand \EOS [0]{\spacefactor3000\relax}%
	\providecommand \BibitemShut  [1]{\csname bibitem#1\endcsname}%
	\let\auto@bib@innerbib\@empty
	\bibitem [{\citenamefont {M\"uhlbauer}\ \emph {et~al.}(2009)\citenamefont
		{M\"uhlbauer}, \citenamefont {Binz}, \citenamefont {Jonietz}, \citenamefont
		{Pfleiderer}, \citenamefont {Rosch}, \citenamefont {Neubauer}, \citenamefont
		{Georgii},\ and\ \citenamefont {Boni}}]{Muhlbauer2009}%
	\BibitemOpen
	\bibfield  {author} {\bibinfo {author} {\bibfnamefont {S.}~\bibnamefont
			{M\"uhlbauer}}, \bibinfo {author} {\bibfnamefont {B.}~\bibnamefont {Binz}},
		\bibinfo {author} {\bibfnamefont {F.}~\bibnamefont {Jonietz}}, \bibinfo
		{author} {\bibfnamefont {C.}~\bibnamefont {Pfleiderer}}, \bibinfo {author}
		{\bibfnamefont {A.}~\bibnamefont {Rosch}}, \bibinfo {author} {\bibfnamefont
			{A.}~\bibnamefont {Neubauer}}, \bibinfo {author} {\bibfnamefont
			{R.}~\bibnamefont {Georgii}},\ and\ \bibinfo {author} {\bibfnamefont
			{P.}~\bibnamefont {Boni}},\ }\bibfield  {title} {\bibinfo {title} {Skyrmion
			lattice in a chiral magnet},\ }\href
	{https://doi.org/10.1126/science.1166767} {\bibfield  {journal} {\bibinfo
			{journal} {Science}\ }\textbf {\bibinfo {volume} {323}},\ \bibinfo {pages}
		{915} (\bibinfo {year} {2009})}\BibitemShut {NoStop}%
	\bibitem [{\citenamefont {Yu}\ \emph {et~al.}(2010)\citenamefont {Yu},
		\citenamefont {Onose}, \citenamefont {Kanazawa}, \citenamefont {Park},
		\citenamefont {Han}, \citenamefont {Matsui}, \citenamefont {Nagaosa},\ and\
		\citenamefont {Tokura}}]{Yu2010}%
	\BibitemOpen
	\bibfield  {author} {\bibinfo {author} {\bibfnamefont {X.}~\bibnamefont
			{Yu}}, \bibinfo {author} {\bibfnamefont {Y.}~\bibnamefont {Onose}}, \bibinfo
		{author} {\bibfnamefont {N.}~\bibnamefont {Kanazawa}}, \bibinfo {author}
		{\bibfnamefont {J.~H.}\ \bibnamefont {Park}}, \bibinfo {author}
		{\bibfnamefont {J.}~\bibnamefont {Han}}, \bibinfo {author} {\bibfnamefont
			{Y.}~\bibnamefont {Matsui}}, \bibinfo {author} {\bibfnamefont
			{N.}~\bibnamefont {Nagaosa}},\ and\ \bibinfo {author} {\bibfnamefont
			{Y.}~\bibnamefont {Tokura}},\ }\bibfield  {title} {\bibinfo {title}
		{Real-space observation of a two-dimensional skyrmion crystal},\ }\href
	{https://doi.org/10.1038/nature09124} {\bibfield  {journal} {\bibinfo
			{journal} {Nature}\ }\textbf {\bibinfo {volume} {465}},\ \bibinfo {pages}
		{901} (\bibinfo {year} {2010})}\BibitemShut {NoStop}%
	\bibitem [{\citenamefont {Heinze}\ \emph {et~al.}(2011)\citenamefont {Heinze},
		\citenamefont {von Bergmann}, \citenamefont {Menzel}, \citenamefont {Brede},
		\citenamefont {Kubetzka}, \citenamefont {Wiesendanger}, \citenamefont
		{Bihlmayer},\ and\ \citenamefont {Bl\"ugel}}]{Heinze2011}%
	\BibitemOpen
	\bibfield  {author} {\bibinfo {author} {\bibfnamefont {S.}~\bibnamefont
			{Heinze}}, \bibinfo {author} {\bibfnamefont {K.}~\bibnamefont {von
				Bergmann}}, \bibinfo {author} {\bibfnamefont {M.}~\bibnamefont {Menzel}},
		\bibinfo {author} {\bibfnamefont {J.}~\bibnamefont {Brede}}, \bibinfo
		{author} {\bibfnamefont {A.}~\bibnamefont {Kubetzka}}, \bibinfo {author}
		{\bibfnamefont {R.}~\bibnamefont {Wiesendanger}}, \bibinfo {author}
		{\bibfnamefont {G.}~\bibnamefont {Bihlmayer}},\ and\ \bibinfo {author}
		{\bibfnamefont {S.}~\bibnamefont {Bl\"ugel}},\ }\bibfield  {title} {\bibinfo
		{title} {{Spontaneous Atomic-Scale Magnetic Skyrmion Lattice in Two
				Dimensions}},\ }\href {https://doi.org/10.1038/nphys2045} {\bibfield
		{journal} {\bibinfo  {journal} {Nat Phys}\ }\textbf {\bibinfo {volume} {7}},\
		\bibinfo {pages} {713} (\bibinfo {year} {2011})}\BibitemShut {NoStop}%
	\bibitem [{\citenamefont {Moreau-Luchaire}\ \emph {et~al.}(2016)\citenamefont
		{Moreau-Luchaire}, \citenamefont {Moutafis}, \citenamefont {Reyren},
		\citenamefont {Sampaio}, \citenamefont {Vaz}, \citenamefont {Horne},
		\citenamefont {Bouzehouane}, \citenamefont {Garcia}, \citenamefont
		{Deranlot}, \citenamefont {Warnicke}, \citenamefont {Wohlh\"{u}ter},
		\citenamefont {George}, \citenamefont {Weigand}, \citenamefont {Raabe},
		\citenamefont {Cros},\ and\ \citenamefont {Fert}}]{Moreau2016}%
	\BibitemOpen
	\bibfield  {author} {\bibinfo {author} {\bibfnamefont {C.}~\bibnamefont
			{Moreau-Luchaire}}, \bibinfo {author} {\bibfnamefont {C.}~\bibnamefont
			{Moutafis}}, \bibinfo {author} {\bibfnamefont {N.}~\bibnamefont {Reyren}},
		\bibinfo {author} {\bibfnamefont {J.}~\bibnamefont {Sampaio}}, \bibinfo
		{author} {\bibfnamefont {C.~A.~F.}\ \bibnamefont {Vaz}}, \bibinfo {author}
		{\bibfnamefont {N.~V.}\ \bibnamefont {Horne}}, \bibinfo {author}
		{\bibfnamefont {K.}~\bibnamefont {Bouzehouane}}, \bibinfo {author}
		{\bibfnamefont {K.}~\bibnamefont {Garcia}}, \bibinfo {author} {\bibfnamefont
			{C.}~\bibnamefont {Deranlot}}, \bibinfo {author} {\bibfnamefont
			{P.}~\bibnamefont {Warnicke}}, \bibinfo {author} {\bibfnamefont
			{P.}~\bibnamefont {Wohlh\"{u}ter}}, \bibinfo {author} {\bibfnamefont {J.-M.}\
			\bibnamefont {George}}, \bibinfo {author} {\bibfnamefont {M.}~\bibnamefont
			{Weigand}}, \bibinfo {author} {\bibfnamefont {J.}~\bibnamefont {Raabe}},
		\bibinfo {author} {\bibfnamefont {V.}~\bibnamefont {Cros}},\ and\ \bibinfo
		{author} {\bibfnamefont {A.}~\bibnamefont {Fert}},\ }\bibfield  {title}
	{\bibinfo {title} {Additive interfacial chiral interaction in multilayers for
			stabilization of small individual skyrmions at room temperature},\ }\href
	{https://doi.org/https://doi.org/10.1038/nnano.2015.313} {\bibfield
		{journal} {\bibinfo  {journal} {Nat. Nanotechnol.}\ }\textbf {\bibinfo
			{volume} {11}},\ \bibinfo {pages} {444} (\bibinfo {year} {2016})}\BibitemShut
	{NoStop}%
	\bibitem [{\citenamefont {Boulle}\ \emph {et~al.}(2016)\citenamefont {Boulle},
		\citenamefont {Vogel}, \citenamefont {Yang}, \citenamefont {Pizzini},
		\citenamefont {de~Souza~Chaves}, \citenamefont {Locatelli}, \citenamefont
		{Mente{\c{s}}}, \citenamefont {Sala}, \citenamefont {Buda-Prejbeanu},
		\citenamefont {Klein}, \citenamefont {Belmeguenai}, \citenamefont
		{Roussign{\'{e}}}, \citenamefont {Stashkevich}, \citenamefont {Ch{\'{e}}rif},
		\citenamefont {Aballe}, \citenamefont {Foerster}, \citenamefont {Chshiev},
		\citenamefont {Auffret}, \citenamefont {Miron},\ and\ \citenamefont
		{Gaudin}}]{Boulle2016}%
	\BibitemOpen
	\bibfield  {author} {\bibinfo {author} {\bibfnamefont {O.}~\bibnamefont
			{Boulle}}, \bibinfo {author} {\bibfnamefont {J.}~\bibnamefont {Vogel}},
		\bibinfo {author} {\bibfnamefont {H.}~\bibnamefont {Yang}}, \bibinfo {author}
		{\bibfnamefont {S.}~\bibnamefont {Pizzini}}, \bibinfo {author} {\bibfnamefont
			{D.}~\bibnamefont {de~Souza~Chaves}}, \bibinfo {author} {\bibfnamefont
			{A.}~\bibnamefont {Locatelli}}, \bibinfo {author} {\bibfnamefont {T.~O.}\
			\bibnamefont {Mente{\c{s}}}}, \bibinfo {author} {\bibfnamefont
			{A.}~\bibnamefont {Sala}}, \bibinfo {author} {\bibfnamefont {L.~D.}\
			\bibnamefont {Buda-Prejbeanu}}, \bibinfo {author} {\bibfnamefont
			{O.}~\bibnamefont {Klein}}, \bibinfo {author} {\bibfnamefont
			{M.}~\bibnamefont {Belmeguenai}}, \bibinfo {author} {\bibfnamefont
			{Y.}~\bibnamefont {Roussign{\'{e}}}}, \bibinfo {author} {\bibfnamefont
			{A.}~\bibnamefont {Stashkevich}}, \bibinfo {author} {\bibfnamefont {S.~M.}\
			\bibnamefont {Ch{\'{e}}rif}}, \bibinfo {author} {\bibfnamefont
			{L.}~\bibnamefont {Aballe}}, \bibinfo {author} {\bibfnamefont
			{M.}~\bibnamefont {Foerster}}, \bibinfo {author} {\bibfnamefont
			{M.}~\bibnamefont {Chshiev}}, \bibinfo {author} {\bibfnamefont
			{S.}~\bibnamefont {Auffret}}, \bibinfo {author} {\bibfnamefont {I.~M.}\
			\bibnamefont {Miron}},\ and\ \bibinfo {author} {\bibfnamefont
			{G.}~\bibnamefont {Gaudin}},\ }\bibfield  {title} {\bibinfo {title}
		{Room-temperature chiral magnetic skyrmions in ultrathin magnetic
			nanostructures},\ }\href
	{https://doi.org/https://doi.org/10.1038/nnano.2015.315} {\bibfield
		{journal} {\bibinfo  {journal} {Nat. Nanotechnol.}\ }\textbf {\bibinfo
			{volume} {11}},\ \bibinfo {pages} {449} (\bibinfo {year} {2016})}\BibitemShut
	{NoStop}%
	\bibitem [{\citenamefont {Woo}\ \emph {et~al.}(2016)\citenamefont {Woo},
		\citenamefont {Litzius}, \citenamefont {Kr\"{u}ger}, \citenamefont {Im},
		\citenamefont {Caretta}, \citenamefont {Richter}, \citenamefont {Mann},
		\citenamefont {Krone}, \citenamefont {Reeve}, \citenamefont {Weigand},
		\citenamefont {Agrawal}, \citenamefont {Lemesh}, \citenamefont {Mawass},
		\citenamefont {Fischer}, \citenamefont {Kl\"{a}ui},\ and\ \citenamefont
		{Beach}}]{Woo2016}%
	\BibitemOpen
	\bibfield  {author} {\bibinfo {author} {\bibfnamefont {S.}~\bibnamefont
			{Woo}}, \bibinfo {author} {\bibfnamefont {K.}~\bibnamefont {Litzius}},
		\bibinfo {author} {\bibfnamefont {B.}~\bibnamefont {Kr\"{u}ger}}, \bibinfo
		{author} {\bibfnamefont {M.-Y.}\ \bibnamefont {Im}}, \bibinfo {author}
		{\bibfnamefont {L.}~\bibnamefont {Caretta}}, \bibinfo {author} {\bibfnamefont
			{K.}~\bibnamefont {Richter}}, \bibinfo {author} {\bibfnamefont
			{M.}~\bibnamefont {Mann}}, \bibinfo {author} {\bibfnamefont {A.}~\bibnamefont
			{Krone}}, \bibinfo {author} {\bibfnamefont {R.~M.}\ \bibnamefont {Reeve}},
		\bibinfo {author} {\bibfnamefont {M.}~\bibnamefont {Weigand}}, \bibinfo
		{author} {\bibfnamefont {P.}~\bibnamefont {Agrawal}}, \bibinfo {author}
		{\bibfnamefont {I.}~\bibnamefont {Lemesh}}, \bibinfo {author} {\bibfnamefont
			{M.-A.}\ \bibnamefont {Mawass}}, \bibinfo {author} {\bibfnamefont
			{P.}~\bibnamefont {Fischer}}, \bibinfo {author} {\bibfnamefont
			{M.}~\bibnamefont {Kl\"{a}ui}},\ and\ \bibinfo {author} {\bibfnamefont
			{G.~S.~D.}\ \bibnamefont {Beach}},\ }\bibfield  {title} {\bibinfo {title}
		{Observation of room-temperature magnetic skyrmions and their current-driven
			dynamics in ultrathin metallic ferromagnets},\ }\href
	{https://doi.org/https://doi.org/10.1038/nmat4593} {\bibfield  {journal}
		{\bibinfo  {journal} {Nat. Mater.}\ }\textbf {\bibinfo {volume} {15}},\
		\bibinfo {pages} {501} (\bibinfo {year} {2016})}\BibitemShut {NoStop}%
	\bibitem [{\citenamefont {Bogdanov}\ and\ \citenamefont
		{Yablonskii}(1989)}]{Bogdanov1989}%
	\BibitemOpen
	\bibfield  {author} {\bibinfo {author} {\bibfnamefont {A.}~\bibnamefont
			{Bogdanov}}\ and\ \bibinfo {author} {\bibfnamefont {D.}~\bibnamefont
			{Yablonskii}},\ }\bibfield  {title} {\bibinfo {title} {{Thermodynamically
				stable ``vortices" in magnetically ordered crystals. The mixed state of
				magnets}},\ }\href@noop {} {\bibfield  {journal} {\bibinfo  {journal} {Sov.
				Phys. JETP}\ }\textbf {\bibinfo {volume} {68}},\ \bibinfo {pages} {101}
		(\bibinfo {year} {1989})}\BibitemShut {NoStop}%
	\bibitem [{\citenamefont {Fert}\ \emph {et~al.}(2017)\citenamefont {Fert},
		\citenamefont {Reyren},\ and\ \citenamefont {Cros}}]{Fert2017}%
	\BibitemOpen
	\bibfield  {author} {\bibinfo {author} {\bibfnamefont {A.}~\bibnamefont
			{Fert}}, \bibinfo {author} {\bibfnamefont {N.}~\bibnamefont {Reyren}},\ and\
		\bibinfo {author} {\bibfnamefont {V.}~\bibnamefont {Cros}},\ }\bibfield
	{title} {\bibinfo {title} {Magnetic skyrmions: advances in physics and
			potential applications},\ }\href
	{https://doi.org/https://doi.org/10.1038/natrevmats.2017.31} {\bibfield
		{journal} {\bibinfo  {journal} {Nat. Rev. Mater.}\ }\textbf {\bibinfo
			{volume} {2}},\ \bibinfo {pages} {17031} (\bibinfo {year}
		{2017})}\BibitemShut {NoStop}%
	\bibitem [{\citenamefont {Back}\ \emph {et~al.}(2020)\citenamefont {Back},
		\citenamefont {Cros}, \citenamefont {Ebert}, \citenamefont {Everschor-Sitte},
		\citenamefont {Fert}, \citenamefont {Garst}, \citenamefont {Ma},
		\citenamefont {Mankovsky}, \citenamefont {Monchesky}, \citenamefont
		{Mostovoy}, \citenamefont {Nagaosa}, \citenamefont {Parkin}, \citenamefont
		{Pfleiderer}, \citenamefont {Reyren}, \citenamefont {Rosch}, \citenamefont
		{Taguchi}, \citenamefont {Tokura}, \citenamefont {von Bergmann},\ and\
		\citenamefont {Zang}}]{Back2020}%
	\BibitemOpen
	\bibfield  {author} {\bibinfo {author} {\bibfnamefont {C.}~\bibnamefont
			{Back}}, \bibinfo {author} {\bibfnamefont {V.}~\bibnamefont {Cros}}, \bibinfo
		{author} {\bibfnamefont {H.}~\bibnamefont {Ebert}}, \bibinfo {author}
		{\bibfnamefont {K.}~\bibnamefont {Everschor-Sitte}}, \bibinfo {author}
		{\bibfnamefont {A.}~\bibnamefont {Fert}}, \bibinfo {author} {\bibfnamefont
			{M.}~\bibnamefont {Garst}}, \bibinfo {author} {\bibfnamefont
			{T.}~\bibnamefont {Ma}}, \bibinfo {author} {\bibfnamefont {S.}~\bibnamefont
			{Mankovsky}}, \bibinfo {author} {\bibfnamefont {T.~L.}\ \bibnamefont
			{Monchesky}}, \bibinfo {author} {\bibfnamefont {M.}~\bibnamefont {Mostovoy}},
		\bibinfo {author} {\bibfnamefont {N.}~\bibnamefont {Nagaosa}}, \bibinfo
		{author} {\bibfnamefont {S.~S.~P.}\ \bibnamefont {Parkin}}, \bibinfo {author}
		{\bibfnamefont {C.}~\bibnamefont {Pfleiderer}}, \bibinfo {author}
		{\bibfnamefont {N.}~\bibnamefont {Reyren}}, \bibinfo {author} {\bibfnamefont
			{A.}~\bibnamefont {Rosch}}, \bibinfo {author} {\bibfnamefont
			{Y.}~\bibnamefont {Taguchi}}, \bibinfo {author} {\bibfnamefont
			{Y.}~\bibnamefont {Tokura}}, \bibinfo {author} {\bibfnamefont
			{K.}~\bibnamefont {von Bergmann}},\ and\ \bibinfo {author} {\bibfnamefont
			{J.}~\bibnamefont {Zang}},\ }\bibfield  {title} {\bibinfo {title} {{The 2020
				skyrmionics roadmap}},\ }\href
	{https://doi.org/https://doi.org/10.1088/1361-6463/ab8418} {\bibfield
		{journal} {\bibinfo  {journal} {J. Phys. D: Appl. Phys.}\ }\textbf {\bibinfo
			{volume} {53}},\ \bibinfo {pages} {363001} (\bibinfo {year}
		{2020})}\BibitemShut {NoStop}%
	\bibitem [{\citenamefont {Dzialoshinskii}(1957)}]{Dzyaloshinskii1957}%
	\BibitemOpen
	\bibfield  {author} {\bibinfo {author} {\bibfnamefont {I.}~\bibnamefont
			{Dzialoshinskii}},\ }\bibfield  {title} {\bibinfo {title} {{A thermodynamic
				theory of ``weak'' ferromagnetism of antiferromagnetics}},\ }\href@noop {}
	{\bibfield  {journal} {\bibinfo  {journal} {Sov. Phys. JETP}\ }\textbf
		{\bibinfo {volume} {5}},\ \bibinfo {pages} {1259–1262} (\bibinfo {year}
		{1957})}\BibitemShut {NoStop}%
	\bibitem [{\citenamefont {Moriya}(1960)}]{Moriya1960}%
	\BibitemOpen
	\bibfield  {author} {\bibinfo {author} {\bibfnamefont {T.}~\bibnamefont
			{Moriya}},\ }\bibfield  {title} {\bibinfo {title} {{New Mechanism of
				Anisotropic Superexchange Interaction}},\ }\href
	{https://doi.org/10.1103/PhysRevLett.4.228} {\bibfield  {journal} {\bibinfo
			{journal} {Phys. Rev. Lett.}\ }\textbf {\bibinfo {volume} {4}},\ \bibinfo
		{pages} {228} (\bibinfo {year} {1960})}\BibitemShut {NoStop}%
	\bibitem [{\citenamefont {Bode}\ \emph {et~al.}(2007)\citenamefont {Bode},
		\citenamefont {Heide}, \citenamefont {Von~Bergmann}, \citenamefont
		{Ferriani}, \citenamefont {Heinze}, \citenamefont {Bihlmayer}, \citenamefont
		{Kubetzka}, \citenamefont {Pietzsch}, \citenamefont {Bl{\"u}gel},\ and\
		\citenamefont {Wiesendanger}}]{Bode2007}%
	\BibitemOpen
	\bibfield  {author} {\bibinfo {author} {\bibfnamefont {M.}~\bibnamefont
			{Bode}}, \bibinfo {author} {\bibfnamefont {M.}~\bibnamefont {Heide}},
		\bibinfo {author} {\bibfnamefont {K.}~\bibnamefont {Von~Bergmann}}, \bibinfo
		{author} {\bibfnamefont {P.}~\bibnamefont {Ferriani}}, \bibinfo {author}
		{\bibfnamefont {S.}~\bibnamefont {Heinze}}, \bibinfo {author} {\bibfnamefont
			{G.}~\bibnamefont {Bihlmayer}}, \bibinfo {author} {\bibfnamefont
			{A.}~\bibnamefont {Kubetzka}}, \bibinfo {author} {\bibfnamefont
			{O.}~\bibnamefont {Pietzsch}}, \bibinfo {author} {\bibfnamefont
			{S.}~\bibnamefont {Bl{\"u}gel}},\ and\ \bibinfo {author} {\bibfnamefont
			{R.}~\bibnamefont {Wiesendanger}},\ }\bibfield  {title} {\bibinfo {title}
		{Chiral magnetic order at surfaces driven by inversion asymmetry},\ }\href
	{https://doi.org/10.1038/nature05802} {\bibfield  {journal} {\bibinfo
			{journal} {Nature}\ }\textbf {\bibinfo {volume} {447}},\ \bibinfo {pages}
		{190} (\bibinfo {year} {2007})}\BibitemShut {NoStop}%
	\bibitem [{\citenamefont {Okubo}\ \emph {et~al.}(2012)\citenamefont {Okubo},
		\citenamefont {Chung},\ and\ \citenamefont {Kawamura}}]{Okubo2012}%
	\BibitemOpen
	\bibfield  {author} {\bibinfo {author} {\bibfnamefont {T.}~\bibnamefont
			{Okubo}}, \bibinfo {author} {\bibfnamefont {S.}~\bibnamefont {Chung}},\ and\
		\bibinfo {author} {\bibfnamefont {H.}~\bibnamefont {Kawamura}},\ }\bibfield
	{title} {\bibinfo {title} {Multiple-$q$ states and the skyrmion lattice of
			the triangular-lattice heisenberg antiferromagnet under magnetic fields},\
	}\href {https://doi.org/10.1103/PhysRevLett.108.017206} {\bibfield  {journal}
		{\bibinfo  {journal} {Phys. Rev. Lett.}\ }\textbf {\bibinfo {volume} {108}},\
		\bibinfo {pages} {017206} (\bibinfo {year} {2012})}\BibitemShut {NoStop}%
	\bibitem [{\citenamefont {Leonov}\ and\ \citenamefont
		{Mostovoy}(2015)}]{Leonov2015}%
	\BibitemOpen
	\bibfield  {author} {\bibinfo {author} {\bibfnamefont {A.~O.}\ \bibnamefont
			{Leonov}}\ and\ \bibinfo {author} {\bibfnamefont {M.}~\bibnamefont
			{Mostovoy}},\ }\bibfield  {title} {\bibinfo {title} {Multiply periodic states
			and isolated skyrmions in an anisotropic frustrated magnet},\ }\href
	{https://doi.org/https://doi.org/10.1038/ncomms9275} {\bibfield  {journal}
		{\bibinfo  {journal} {Nat. Commun.}\ }\textbf {\bibinfo {volume} {6}},\
		\bibinfo {pages} {8275} (\bibinfo {year} {2015})}\BibitemShut {NoStop}%
	\bibitem [{\citenamefont {von Malottki}\ \emph {et~al.}(2017)\citenamefont {von
			Malottki}, \citenamefont {Dupé}, \citenamefont {Bessarab}, \citenamefont
		{Delin},\ and\ \citenamefont {Heinze}}]{Malottki2017}%
	\BibitemOpen
	\bibfield  {author} {\bibinfo {author} {\bibfnamefont {S.}~\bibnamefont {von
				Malottki}}, \bibinfo {author} {\bibfnamefont {B.}~\bibnamefont {Dupé}},
		\bibinfo {author} {\bibfnamefont {P.}~\bibnamefont {Bessarab}}, \bibinfo
		{author} {\bibfnamefont {A.}~\bibnamefont {Delin}},\ and\ \bibinfo {author}
		{\bibfnamefont {S.}~\bibnamefont {Heinze}},\ }\bibfield  {title} {\bibinfo
		{title} {{Enhanced skyrmion stability due to exchange frustration}},\ }\href
	{https://doi.org/10.1038/s41598-017-12525-x} {\bibfield  {journal} {\bibinfo
			{journal} {Sci. Rep.}\ }\textbf {\bibinfo {volume} {7}},\ \bibinfo {pages}
		{12299} (\bibinfo {year} {2017})}\BibitemShut {NoStop}%
	\bibitem [{\citenamefont {Paul}\ \emph {et~al.}(2020)\citenamefont {Paul},
		\citenamefont {Haldar}, \citenamefont {Malottki},\ and\ \citenamefont
		{Heinze}}]{Paul2020}%
	\BibitemOpen
	\bibfield  {author} {\bibinfo {author} {\bibfnamefont {S.}~\bibnamefont
			{Paul}}, \bibinfo {author} {\bibfnamefont {S.}~\bibnamefont {Haldar}},
		\bibinfo {author} {\bibfnamefont {S.}~\bibnamefont {Malottki}},\ and\
		\bibinfo {author} {\bibfnamefont {S.}~\bibnamefont {Heinze}},\ }\bibfield
	{title} {\bibinfo {title} {Role of higher-order exchange interactions for
			skyrmion stability},\ }\href {https://doi.org/10.1038/s41467-020-18473-x}
	{\bibfield  {journal} {\bibinfo  {journal} {Nat. Commun.}\ }\textbf {\bibinfo
			{volume} {11}},\ \bibinfo {pages} {4756} (\bibinfo {year}
		{2020})}\BibitemShut {NoStop}%
	\bibitem [{\citenamefont {Kurz}\ \emph {et~al.}(2001)\citenamefont {Kurz},
		\citenamefont {Bihlmayer}, \citenamefont {Hirai},\ and\ \citenamefont
		{Bl\"ugel}}]{Kurz2001}%
	\BibitemOpen
	\bibfield  {author} {\bibinfo {author} {\bibfnamefont {P.}~\bibnamefont
			{Kurz}}, \bibinfo {author} {\bibfnamefont {G.}~\bibnamefont {Bihlmayer}},
		\bibinfo {author} {\bibfnamefont {K.}~\bibnamefont {Hirai}},\ and\ \bibinfo
		{author} {\bibfnamefont {S.}~\bibnamefont {Bl\"ugel}},\ }\bibfield  {title}
	{\bibinfo {title} {{Three-Dimensional Spin Structure on a Two-Dimensional
				Lattice: Mn $/$Cu(111)}},\ }\href
	{https://doi.org/10.1103/PhysRevLett.86.1106} {\bibfield  {journal} {\bibinfo
			{journal} {Phys. Rev. Lett.}\ }\textbf {\bibinfo {volume} {86}},\ \bibinfo
		{pages} {1106} (\bibinfo {year} {2001})}\BibitemShut {NoStop}%
	\bibitem [{\citenamefont {Kr\"onlein}\ \emph {et~al.}(2018)\citenamefont
		{Kr\"onlein}, \citenamefont {Schmitt}, \citenamefont {Hoffmann},
		\citenamefont {Kemmer}, \citenamefont {Seubert}, \citenamefont {Vogt},
		\citenamefont {K\"uspert}, \citenamefont {B\"ohme}, \citenamefont {Alonazi},
		\citenamefont {K\"ugel}, \citenamefont {Albrithen}, \citenamefont {Bode},
		\citenamefont {Bihlmayer},\ and\ \citenamefont {Bl\"ugel}}]{Kroenlein2018}%
	\BibitemOpen
	\bibfield  {author} {\bibinfo {author} {\bibfnamefont {A.}~\bibnamefont
			{Kr\"onlein}}, \bibinfo {author} {\bibfnamefont {M.}~\bibnamefont {Schmitt}},
		\bibinfo {author} {\bibfnamefont {M.}~\bibnamefont {Hoffmann}}, \bibinfo
		{author} {\bibfnamefont {J.}~\bibnamefont {Kemmer}}, \bibinfo {author}
		{\bibfnamefont {N.}~\bibnamefont {Seubert}}, \bibinfo {author} {\bibfnamefont
			{M.}~\bibnamefont {Vogt}}, \bibinfo {author} {\bibfnamefont {J.}~\bibnamefont
			{K\"uspert}}, \bibinfo {author} {\bibfnamefont {M.}~\bibnamefont {B\"ohme}},
		\bibinfo {author} {\bibfnamefont {B.}~\bibnamefont {Alonazi}}, \bibinfo
		{author} {\bibfnamefont {J.}~\bibnamefont {K\"ugel}}, \bibinfo {author}
		{\bibfnamefont {H.~A.}\ \bibnamefont {Albrithen}}, \bibinfo {author}
		{\bibfnamefont {M.}~\bibnamefont {Bode}}, \bibinfo {author} {\bibfnamefont
			{G.}~\bibnamefont {Bihlmayer}},\ and\ \bibinfo {author} {\bibfnamefont
			{S.}~\bibnamefont {Bl\"ugel}},\ }\bibfield  {title} {\bibinfo {title}
		{{Magnetic Ground State Stabilized by Three-Site Interactions:
				$\mathrm{Fe}/\mathrm{Rh}(111)$}},\ }\href
	{https://doi.org/10.1103/PhysRevLett.120.207202} {\bibfield  {journal}
		{\bibinfo  {journal} {Phys. Rev. Lett.}\ }\textbf {\bibinfo {volume} {120}},\
		\bibinfo {pages} {207202} (\bibinfo {year} {2018})}\BibitemShut {NoStop}%
	\bibitem [{\citenamefont {Romming}\ \emph {et~al.}(2018)\citenamefont
		{Romming}, \citenamefont {Pralow}, \citenamefont {Kubetzka}, \citenamefont
		{Hoffmann}, \citenamefont {von Malottki}, \citenamefont {Meyer},
		\citenamefont {Dup\'e}, \citenamefont {Wiesendanger}, \citenamefont {von
			Bergmann},\ and\ \citenamefont {Heinze}}]{Romming2018}%
	\BibitemOpen
	\bibfield  {author} {\bibinfo {author} {\bibfnamefont {N.}~\bibnamefont
			{Romming}}, \bibinfo {author} {\bibfnamefont {H.}~\bibnamefont {Pralow}},
		\bibinfo {author} {\bibfnamefont {A.}~\bibnamefont {Kubetzka}}, \bibinfo
		{author} {\bibfnamefont {M.}~\bibnamefont {Hoffmann}}, \bibinfo {author}
		{\bibfnamefont {S.}~\bibnamefont {von Malottki}}, \bibinfo {author}
		{\bibfnamefont {S.}~\bibnamefont {Meyer}}, \bibinfo {author} {\bibfnamefont
			{B.}~\bibnamefont {Dup\'e}}, \bibinfo {author} {\bibfnamefont
			{R.}~\bibnamefont {Wiesendanger}}, \bibinfo {author} {\bibfnamefont
			{K.}~\bibnamefont {von Bergmann}},\ and\ \bibinfo {author} {\bibfnamefont
			{S.}~\bibnamefont {Heinze}},\ }\bibfield  {title} {\bibinfo {title}
		{{Competition of Dzyaloshinskii-Moriya and Higher-Order Exchange Interactions
				in $\mathrm{Rh}/\mathrm{Fe}$ Atomic Bilayers on Ir(111)}},\ }\href
	{https://doi.org/10.1103/PhysRevLett.120.207201} {\bibfield  {journal}
		{\bibinfo  {journal} {Phys. Rev. Lett.}\ }\textbf {\bibinfo {volume} {120}},\
		\bibinfo {pages} {207201} (\bibinfo {year} {2018})}\BibitemShut {NoStop}%
	\bibitem [{\citenamefont {Spethmann}\ \emph {et~al.}(2020)\citenamefont
		{Spethmann}, \citenamefont {Meyer}, \citenamefont {von Bergmann},
		\citenamefont {Wiesendanger}, \citenamefont {Heinze},\ and\ \citenamefont
		{Kubetzka}}]{Spethmann2020}%
	\BibitemOpen
	\bibfield  {author} {\bibinfo {author} {\bibfnamefont {J.}~\bibnamefont
			{Spethmann}}, \bibinfo {author} {\bibfnamefont {S.}~\bibnamefont {Meyer}},
		\bibinfo {author} {\bibfnamefont {K.}~\bibnamefont {von Bergmann}}, \bibinfo
		{author} {\bibfnamefont {R.}~\bibnamefont {Wiesendanger}}, \bibinfo {author}
		{\bibfnamefont {S.}~\bibnamefont {Heinze}},\ and\ \bibinfo {author}
		{\bibfnamefont {A.}~\bibnamefont {Kubetzka}},\ }\bibfield  {title} {\bibinfo
		{title} {{Discovery of magnetic single- and triple-Q states in
				Mn/Re(0001)}},\ }\href {https://doi.org/10.1103/PhysRevLett.124.227203}
	{\bibfield  {journal} {\bibinfo  {journal} {Phys. Rev. Lett.}\ }\textbf
		{\bibinfo {volume} {124}},\ \bibinfo {pages} {227203} (\bibinfo {year}
		{2020})}\BibitemShut {NoStop}%
	\bibitem [{\citenamefont {Li}\ \emph {et~al.}(2020)\citenamefont {Li},
		\citenamefont {Paul}, \citenamefont {von Bergmann}, \citenamefont {Heinze},\
		and\ \citenamefont {Wiesendanger}}]{Li2020}%
	\BibitemOpen
	\bibfield  {author} {\bibinfo {author} {\bibfnamefont {W.}~\bibnamefont
			{Li}}, \bibinfo {author} {\bibfnamefont {S.}~\bibnamefont {Paul}}, \bibinfo
		{author} {\bibfnamefont {K.}~\bibnamefont {von Bergmann}}, \bibinfo {author}
		{\bibfnamefont {S.}~\bibnamefont {Heinze}},\ and\ \bibinfo {author}
		{\bibfnamefont {R.}~\bibnamefont {Wiesendanger}},\ }\bibfield  {title}
	{\bibinfo {title} {{Stacking-Dependent Spin Interactions in
				$\mathrm{Pd}/\mathrm{Fe}$ Bilayers on Re(0001)}},\ }\href
	{https://doi.org/10.1103/PhysRevLett.125.227205} {\bibfield  {journal}
		{\bibinfo  {journal} {Phys. Rev. Lett.}\ }\textbf {\bibinfo {volume} {125}},\
		\bibinfo {pages} {227205} (\bibinfo {year} {2020})}\BibitemShut {NoStop}%
	\bibitem [{\citenamefont {Haldar}\ \emph {et~al.}(2021)\citenamefont {Haldar},
		\citenamefont {Meyer}, \citenamefont {Kubetzka},\ and\ \citenamefont
		{Heinze}}]{Haldar2021}%
	\BibitemOpen
	\bibfield  {author} {\bibinfo {author} {\bibfnamefont {S.}~\bibnamefont
			{Haldar}}, \bibinfo {author} {\bibfnamefont {S.}~\bibnamefont {Meyer}},
		\bibinfo {author} {\bibfnamefont {A.}~\bibnamefont {Kubetzka}},\ and\
		\bibinfo {author} {\bibfnamefont {S.}~\bibnamefont {Heinze}},\ }\bibfield
	{title} {\bibinfo {title} {{Distorted 3Q state driven by topological-chiral
				magnetic interactions}},\ }\href
	{https://doi.org/10.1103/PhysRevB.104.L180404} {\bibfield  {journal}
		{\bibinfo  {journal} {Phys. Rev. B}\ }\textbf {\bibinfo {volume} {104}},\
		\bibinfo {pages} {L180404} (\bibinfo {year} {2021})}\BibitemShut {NoStop}%
	\bibitem [{\citenamefont {Gutzeit}\ \emph {et~al.}(2021)\citenamefont
		{Gutzeit}, \citenamefont {Haldar}, \citenamefont {Meyer},\ and\ \citenamefont
		{Heinze}}]{Gutzeit2021}%
	\BibitemOpen
	\bibfield  {author} {\bibinfo {author} {\bibfnamefont {M.}~\bibnamefont
			{Gutzeit}}, \bibinfo {author} {\bibfnamefont {S.}~\bibnamefont {Haldar}},
		\bibinfo {author} {\bibfnamefont {S.}~\bibnamefont {Meyer}},\ and\ \bibinfo
		{author} {\bibfnamefont {S.}~\bibnamefont {Heinze}},\ }\bibfield  {title}
	{\bibinfo {title} {Trends of higher-order exchange interactions in transition
			metal trilayers},\ }\href {https://doi.org/10.1103/PhysRevB.104.024420}
	{\bibfield  {journal} {\bibinfo  {journal} {Phys. Rev. B}\ }\textbf {\bibinfo
			{volume} {104}},\ \bibinfo {pages} {024420} (\bibinfo {year}
		{2021})}\BibitemShut {NoStop}%
	\bibitem [{\citenamefont {Hoffmann}\ and\ \citenamefont
		{Bl\"ugel}(2020)}]{Hoffmann2020}%
	\BibitemOpen
	\bibfield  {author} {\bibinfo {author} {\bibfnamefont {M.}~\bibnamefont
			{Hoffmann}}\ and\ \bibinfo {author} {\bibfnamefont {S.}~\bibnamefont
			{Bl\"ugel}},\ }\bibfield  {title} {\bibinfo {title} {{Systematic derivation
				of realistic spin models for beyond-Heisenberg solids}},\ }\href
	{https://doi.org/10.1103/PhysRevB.101.024418} {\bibfield  {journal} {\bibinfo
			{journal} {Phys. Rev. B}\ }\textbf {\bibinfo {volume} {101}},\ \bibinfo
		{pages} {024418} (\bibinfo {year} {2020})}\BibitemShut {NoStop}%
	\bibitem [{\citenamefont {Gutzeit}\ \emph {et~al.}(2022)\citenamefont
		{Gutzeit}, \citenamefont {Kubetzka}, \citenamefont {Haldar}, \citenamefont
		{Pralow}, \citenamefont {Goerzen}, \citenamefont {Wiesendanger},
		\citenamefont {Heinze},\ and\ \citenamefont {von Bergmann}}]{Gutzeit2022}%
	\BibitemOpen
	\bibfield  {author} {\bibinfo {author} {\bibfnamefont {M.}~\bibnamefont
			{Gutzeit}}, \bibinfo {author} {\bibfnamefont {A.}~\bibnamefont {Kubetzka}},
		\bibinfo {author} {\bibfnamefont {S.}~\bibnamefont {Haldar}}, \bibinfo
		{author} {\bibfnamefont {H.}~\bibnamefont {Pralow}}, \bibinfo {author}
		{\bibfnamefont {M.}~\bibnamefont {Goerzen}}, \bibinfo {author} {\bibfnamefont
			{R.}~\bibnamefont {Wiesendanger}}, \bibinfo {author} {\bibfnamefont
			{S.}~\bibnamefont {Heinze}},\ and\ \bibinfo {author} {\bibfnamefont
			{K.}~\bibnamefont {von Bergmann}},\ }\bibfield  {title} {\bibinfo {title}
		{Nano-scale collinear multi-q states driven by higher-order interactions},\
	}\href {https://doi.org/https://doi.org/10.1038/s41467-022-33383-w}
	{\bibfield  {journal} {\bibinfo  {journal} {Nat. Commun.}\ }\textbf {\bibinfo
			{volume} {13}},\ \bibinfo {pages} {5764} (\bibinfo {year}
		{2022})}\BibitemShut {NoStop}%
	\bibitem [{\citenamefont {von Bergmann}\ \emph {et~al.}(2015)\citenamefont {von
			Bergmann}, \citenamefont {Menzel}, \citenamefont {Kubetzka},\ and\
		\citenamefont {Wiesendanger}}]{Bergmann2015}%
	\BibitemOpen
	\bibfield  {author} {\bibinfo {author} {\bibfnamefont {K.}~\bibnamefont {von
				Bergmann}}, \bibinfo {author} {\bibfnamefont {M.}~\bibnamefont {Menzel}},
		\bibinfo {author} {\bibfnamefont {A.}~\bibnamefont {Kubetzka}},\ and\
		\bibinfo {author} {\bibfnamefont {R.}~\bibnamefont {Wiesendanger}},\
	}\bibfield  {title} {\bibinfo {title} {Influence of the local atom
			configuration on a hexagonal skyrmion lattice},\ }\href
	{https://doi.org/10.1021/acs.nanolett.5b00506} {\bibfield  {journal}
		{\bibinfo  {journal} {Nano Lett.}\ }\textbf {\bibinfo {volume} {15}},\
		\bibinfo {pages} {3280} (\bibinfo {year} {2015})}\BibitemShut {NoStop}%
	\bibitem [{sup()}]{supplmat}%
	\BibitemOpen
	\href@noop {} {}\bibinfo {note} {See Supplemental Material for computational
		details, additional energy dispersions, calculation of higher order
		interactions, and SP-STM simulations}\BibitemShut {NoStop}%
	\bibitem [{FLE()}]{FLEUR}%
	\BibitemOpen
	\href@noop {} {}\bibinfo {howpublished} {See
		\url{https://www.flapw.de}}\BibitemShut {NoStop}%
	\bibitem [{\citenamefont {von Bergmann}\ \emph {et~al.}(2006)\citenamefont {von
			Bergmann}, \citenamefont {Heinze}, \citenamefont {Bode}, \citenamefont
		{Vedmedenko}, \citenamefont {Bihlmayer}, \citenamefont {Bl\"ugel},\ and\
		\citenamefont {Wiesendanger}}]{vonBergmann2006}%
	\BibitemOpen
	\bibfield  {author} {\bibinfo {author} {\bibfnamefont {K.}~\bibnamefont {von
				Bergmann}}, \bibinfo {author} {\bibfnamefont {S.}~\bibnamefont {Heinze}},
		\bibinfo {author} {\bibfnamefont {M.}~\bibnamefont {Bode}}, \bibinfo {author}
		{\bibfnamefont {E.~Y.}\ \bibnamefont {Vedmedenko}}, \bibinfo {author}
		{\bibfnamefont {G.}~\bibnamefont {Bihlmayer}}, \bibinfo {author}
		{\bibfnamefont {S.}~\bibnamefont {Bl\"ugel}},\ and\ \bibinfo {author}
		{\bibfnamefont {R.}~\bibnamefont {Wiesendanger}},\ }\bibfield  {title}
	{\bibinfo {title} {Observation of a complex nanoscale magnetic structure in a
			hexagonal {F}e monolayer},\ }\href
	{https://doi.org/10.1103/PhysRevLett.96.167203} {\bibfield  {journal}
		{\bibinfo  {journal} {Phys. Rev. Lett.}\ }\textbf {\bibinfo {volume} {96}},\
		\bibinfo {pages} {167203} (\bibinfo {year} {2006})}\BibitemShut {NoStop}%
	\bibitem [{Note1()}]{Note1}%
	\BibitemOpen
	\bibinfo {note} {Note, that the 12-SkX state can also occur as an {\protect
			\it on-top-state} in which the point of constructive interference of the
		three spin spirals is placed on top of a Fe atom \cite {Bergmann2015}. In
		contrast, for the 12-SkX state of Fig.~2(c) this point is on a hollow site.
		Within our DFT calculations, the {\protect \it on-top} and {\protect \it
			hollow} 12-SkX states are energetically nearly degenerate (see Supplemental
		Material \cite {supplmat}).}\BibitemShut {Stop}%
	\bibitem [{Note2()}]{Note2}%
	\BibitemOpen
	\bibinfo {note} {Note, that the energies of all spin spiral (1Q) states in
		the spin model are equal to the DFT values.}\BibitemShut {Stop}%
	\bibitem [{Note3()}]{Note3}%
	\BibitemOpen
	\bibinfo {note} {{ \protect Note that $K_1$ is of similar
			magnitude for both Fe stackings while $B_1$ and $Y_1$ are much smaller for
			hcp-Fe (cf.~Table \ref {Table1}).}}\BibitemShut {Stop}%
\end{thebibliography}
\end{document}